\documentclass[twocolumn,pra,showpacs,flushbottom,english]{revtex4}
\usepackage{graphicx}
\usepackage{latexsym}
\usepackage{amsmath}
\usepackage{amssymb}
\usepackage{times}

\begin{document}

\title{ Master equations for correlated quantum channels}
\author{V. Giovannetti$^{1}$ and G. M. Palma$^{2}$}

\affiliation{$^1$ NEST, Scuola Normale Superiore and Istituto Nanoscienze-CNR, 
piazza dei Cavalieri 7, I-56126 Pisa, Italy 
\\ $^2$NEST  Istituto Nanoscienze-CNR and Dipartimento di Fisica,
Universita' degli Studi di Palermo, via Archirafi 36,
I-90123 Palermo, Italy}

\begin{abstract}
We derive the general form of a master equation describing the interaction of an arbitrary  multipartite quantum system, consisting of a set of subsystems, with an environment, consisting of a large number of sub-envirobments. Each subsystem ``collides" with the same sequence of sub-environments which, in between the collisions, evolve according to a map that mimics relaxations effects. No  assumption is made on the specific nature of neither the system nor the environment. 
 In the weak coupling regime, we show that the collisional model produces a correlated Markovian evolution for the joint density matrix of the multipartite system. The associated Linblad super-operator  contains pairwise terms describing cross correlation between the different subsystems.  
\end{abstract}

\maketitle

In the study of  the open dynamics of a multipartite quantum system  ${\cal S}$ the simplifying assumption that each subsystem interacts with its own local environment is frequently made. In quantum communication~\cite{BENSHOR} where ${\cal S}$ is identified with the set of information carriers employed in the signaling process, this is equivalent to saying that a given communication channel is \emph{memoryless} i.e. that  it acts independently on each separate 
carrier. 
 In recent years, however, the study of \emph{correlated} channels - sometimes called also channels with memory -  has shown that interesting new features emerge when one makes the realistic assumption that the action of the noise tampering with the communication line is  correlated   over consecutive carriers (e.g. see~\cite{ MPV, BM,KW,VJP,PV,LUPO,DARRIGO} and references therein).  Such  correlations have been  phenomenologically described in terms of a Markov chain which gives the joint probability distribution of the local Kraus operators acting on the elements of ${\cal S}$~\cite{MPV}. Alternatively they have been effectively represented in terms of local interactions of the carriers with a common multipartite environment  which is originally prepared into a correlated 
(possibly entangled) initial state~\cite{PV}, or 
with a structured environment composed by local and global components~\cite{BM,KW,VJP}.

The aim of the present paper  is to provide a continuous time description of correlated quantum channels in terms of 
a joint Master Equation (ME)~\cite{LIND,PET} for  ${\cal S}$.  
This will lead us to identify the structure of the Lindblad generators  which are responsible for the arising of  specific correlations among the carriers.
We remind that 
determining if a given quantum transformation is compatible with  a Lindblad structure is in general a computationally hard problem~\cite{NONMARK}.  Also we notice that
 a Lindbladian structure for the global system  ${\cal S}$ in general may introduce non-Markovian
 elements  in the dynamics of the subsystems that compose it, which also 
  are far from trivial to characterize, e.g. see Refs.~\cite{BRVC}.
To bypass such difficulties  in our analysis we will thus  adopt a rather pragmatic approach, 
deriving   the dynamical evolution of ${\cal S}$ from a collisional model~\cite{SZS,ZSB} in which dissipative effects originate from a sequence of 
weak but frequent interactions with a collection of uncorrelated  particles which mimic  the system environment. 
\begin{figure}[t]
\begin{center}
\includegraphics[width=130pt]{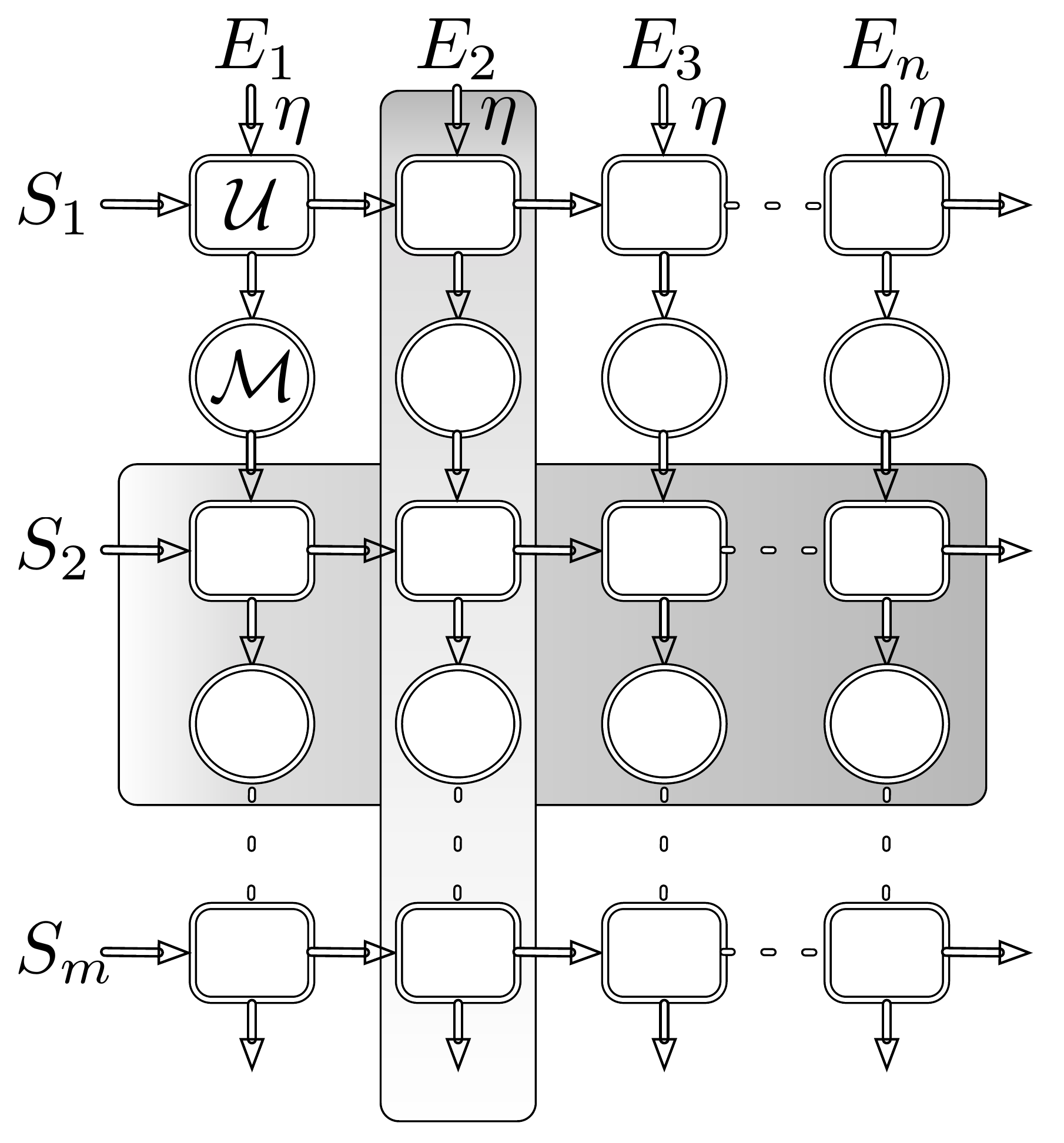} 
\caption{Schematic of the process. The horizontal lines describe an ordered set of  carriers  $S_1, S_2, \cdots$ which interact with an ordered set of (possibly infinite) identical local sub-environments  $E_1, E_2, \cdots$ via local unitaries $\mathcal{U}_{S_mE_n}$. Between collisions each sub-environment evolves according to a map $\mathcal{M}$. The overall dynamics can be described as a ordered sequence of row or of column super-operators  (visualized by the  rectangular sets in the figure).}
\label{fig1}
\end{center}
\end{figure}
Consider hence a multipartite quantum system ${\cal S}$, consisting of $M$  - not necessarily identical - ordered subsystems $S_1,  S_2, \cdots, S_M$. 
In what follows each subsystem is supposed to interact with a multipartite environment ${\cal E}$ consisting of a large number of sub-environments 
$E_1, E_2, \cdots$ via an ordered sequence of pairwise interactions
 (for a pictorial representation see Fig.~\ref{fig1}).
 As in~\cite{SZS,ZSB} the pairwise collision between the subsystem $S_m$ and the sub-envirinment $E_n$ is described by a local unitary $U_{S_mE_n} =\exp[ -i g H_{S_mE_n} \Delta t]$ characterized by a collision time $\Delta t$ and by the intensity parameter $g$, and  generated by the Hamiltonian coupling 
$H_{S_mE_n}$ which (without loss of generality) we write as 
\begin{eqnarray}\label{ggrss}
 H_{S_mE_n} := \sum_\ell \; A^{(\ell)}_{S_m}  \otimes B^{(\ell)}_{E_n} \;,
 \end{eqnarray} 
 with $A^{(\ell)}_{S}, B^{(\ell)}_{E}\neq 0$ Hermitian.
Accordingly  the $m$-th carrier  interacts with  the first $n$ elements  of the environment ${\cal E}$  through the joint unitary
\begin{eqnarray}\label{hh2}
U_{S_m {\cal E}}^{(n)}
 :=  U_{S_m,E_n}\;U_{S_m,E_{n-1}} \;\cdots \; U_{S_m,E_2} \; U_{S_m,E_1}\;,
\end{eqnarray} 
(the presence of a local free Hamiltonian evolution operating  between the collisions can be included in the model by passing into the interaction picture representation
and replacing $A^{(\ell)}_{S_m}$ with the corresponding evolved operators). 
Finally to account for the internal dynamics of the environment, we assume that between two consecutive collisions each sub-environment evolves according to a completely, positive,
 trace preserving (CPT) map $\mathcal{M}$.

Consider then the case where the $M$ subsystems of 
${\cal S}$ are  initially in a (possibly correlated) state $\rho(0)$ while the  
sub-environments of ${\cal E}$ are all prepared into the same input state $\eta$ 
(which, as in Ref.~\cite{SZS}, represents some equilibrium state of the particles of the reservoir).
For the sake of simplicity in the following we will work under the hypothesis that
\begin{eqnarray}~\label{ASSUMPTION1}
\langle B_E^{(\ell)} {\cal M}^{m} (\eta) \rangle_E = 0\qquad  \forall \ell, m \;,
\end{eqnarray} 
where we use the symbol $\langle \cdots \rangle_{X}$ to represent the trace over the system ${X}$, and 
 ${\cal M}^{m}$ to represent the channel obtained by applying $m$ times the map ${\cal M}$.
The assumption~(\ref{ASSUMPTION1})  allows us to rigorously define the 
continuous limit of the model. It  is worth noticing however that it does  not imply any loss of generality as it can always be enforced by moving into 
an interaction representation with respect to a rescaled
local Hamiltonian for the system ${\cal S}$.

After the interactions with the first $n$ element of ${\cal E}$ the global state $R(n)$ 
of the system and of the environment is obtained from the initial state $\rho(0) \otimes\eta^{\otimes n}$ as
$R(n) = {\cal W}^{(n,M)} (\rho(0) \otimes\eta^{\otimes n})$,
where ${\cal W}^{(n,M)}$ is the super-operator which describes
the collisions and the free evolutions of ${\cal E}$.
As schematically shown in Fig.~\ref{fig1},  
it can be expressed as a composition of {\em row} super-operators
stack in series one on top of the other
\begin{eqnarray}\label{ROW}
{\cal W}^{(n,M)}  =  {\cal R}_{S_M, {\cal E}} \circ  {\cal R}_{S_{M-1}, {\cal E}} \circ  \cdots\circ   {\cal R}_{S_2, {\cal E}} \circ  {\cal R}_{S_1, {\cal E}} \;,
\end{eqnarray}
 where
 ${\cal R}_{S_m,{\cal E}} := {\cal M}^{\otimes n} \;\circ\; {\cal U}_{S_{m},{\cal E}}^{(n)}$.
 Here given, a unitary transformation $U$, we define ${\cal U}(\cdots)  = U (\cdots) U^\dag$. Also
 we use the symbol ``$\circ$"  to represent the composition of super-operators and 
 ${\cal M}^{\otimes n}$ to represent ${\cal M}_{E_1} \circ \cdots \circ {\cal M}_{E_n}$, ${\cal M}_{E_j}$ being the map ${\cal M}$ operating on the $j$-th element ${E}_j$ of ${\cal E}$. 
 The transformation ${\cal R}_{S_m,{\cal E}}$ describes the evolution of $S_m$ in its interaction with ${\cal E}$ plus the subsequent free evolution of the latter induced by  the maps~$\cal M$.  
 Alternatively, exploiting the fact 
 that for $m'\neq m$, $n'\neq n$ the operators $U_{S_m, E_n}$ and $U_{S_{m'}, E_{n'}}$ commute, 
   ${\cal W}^{(n,M)}$ can also be expressed in terms of  {\em column} super-operators concatenated in series as follows:
 \begin{eqnarray}\label{colu}
{\cal W}^{(n,M)}  =  {\cal C}_{{\cal S}, E_n} \circ {\cal C}_{{\cal S}, E_{n-1}} \circ \cdots \circ {\cal C}_{{\cal S}, E_2} \circ {\cal C}_{{\cal S}, E_1} \;,
 \end{eqnarray} 
 where for all $j =1, \cdots, n$,
 \begin{eqnarray}
 {\cal C}_{{\cal S}, E_j} &:=& {\cal M}_{ E_j} \circ {\cal U}_{S_{M},E_j}  \circ \cdots
   \circ {\cal M}_{ E_j}
 \circ {\cal U}_{S_{1},E_j}\;. \label{COLUMN}
 \end{eqnarray} 
Thanks to Eq.~(\ref{colu}) we can now write the following recursive expression for $R(n)$:
\begin{eqnarray}
R(n+1)&=&{\cal C}_{{\cal S}, E_{n+1}}  (R(n)\otimes \eta)\;. \label{rec}
\end{eqnarray} 
\section{The Master Equation} 

 For a particular  class of interaction unitaries, the Authors of~\cite{ZSB} have shown 
that the collision model leads to a dynamics which can be described by a 
Lindblad super-operator via direct integration of the equation of motion.  Here we introduce an alternative approach which allows one 
to derive a ME  for the reduced dynamics of the many-body system $\mathcal{S}$ in our generalized multipartite collision model.  The details of the derivation can be found in the Appendix. 
We simply assume
 a weak coupling regime where we take a proper expansion with respect to the parameters $g$ and $\Delta t$ which quantifies the intensity and the duration of the single events. 
 In particular we  work in the regime in which  $g\Delta t$ is a small quantity and expand the dynamical equation~(\ref{rec})
  up to  ${\cal O}\big((g\Delta t)^2\big)$, i.e. 
 \begin{eqnarray}
R(n+1)&=&
\big[ {\cal I}_{{\cal S}, E_{n+1}}+  {\cal C}^{\prime}_{{\cal S}, E_{n+1}} g \Delta t \\ \nonumber 
&+&  {\cal C}^{\prime\prime}_{{\cal S}, E_{n+1}} (g\Delta t)^2  \big]  (R(n)\otimes \eta)  + {\cal O}\big((g\Delta t)^3\big)
\;,\nonumber  \label{EQperR}
\end{eqnarray} 
where ${\cal I}_{{\cal S}, E_{n+1}}$ is the identity superoperator while ${\cal C}^{\prime}_{{\cal S}, E_{n+1}}$ and ${\cal C}^{\prime\prime}_{{\cal S}, E_{n+1}}$ are the first and second expansion terms in $g\Delta t$  of 
the superoperator  ${\cal C}_{{\cal S}, E_{n+1}}$, respectively. Tracing over the
 degree of freedom of the environment the resulting equation defines the incremental evolution of the density matrix $\rho(n) :=\langle R(n)\rangle_{\cal E}$ of ${\cal S}$ when passing from the $n$-th  to the  $(n+1)$-th collision. The continuos limit is finally taken by sending $\Delta t$ to zero while $g$ and $n$ explode in such a way that  $n \Delta t$  and $g^2 \Delta t$ remains finite,  i.e.  
\begin{eqnarray} 
\lim_{\Delta t  \rightarrow 0^+} n \; \Delta t = t < \infty \;,\quad 
\lim_{\Delta t  \rightarrow 0^+} g^2 \Delta t = \gamma < \infty \;.
\label{defgamma} 
\end{eqnarray} 
Notice that while the first condition  is necessary to properly define the axis of time, the second 
is needed to guarantee that ${\cal S}$ fills the interactions with ${\cal E}$. Indeed one easily verifies   that 
 the linear terms in $g$ do not enter in the dynamical evolution of $\cal S$ since 
 $\langle {\cal C}^{(1)}_{{\cal S}, E_{n+1}}  (R(n)\otimes \eta)\rangle_{\cal E}=0$ due to the assumption~(\ref{ASSUMPTION1}). 

Defining hence  $\rho(t)= \lim_{\Delta t \rightarrow 0^+} \;  \rho(n)$ the reduced density matrix of ${\cal S}$ at time $t$,  and 
$\dot{\rho}(t):= \lim_{\Delta t \rightarrow 0^+} \; \tfrac{\rho(n+1) -\rho(n)}{\Delta t}$  its time derivative,
from Eq.~(\ref{EQperR}) we get the following ME: 
 \begin{eqnarray}
{\dot \rho}(t) = 
 \sum_{m=1}^M {\cal L}_m (\rho(t) )     
 + \sum_{m'>m} {\cal D}^{(\rightarrow)}_{m,m'} (\rho(t) ) \;. \label{fin}
 \end{eqnarray} 
 This is mathematically equivalent to the standard derivation of a Markovian ME
for a system inetracting with a large environment, in which one assumes that  the overall system-environment density operator 
at any given time $t$ of the evolution factorizes as in $\rho(t) \otimes \eta$ where  $\eta$ is the environment density operator. The two scenarios are however different. In the standard case the reason for which the environment state is unchanged is because it is big. In our scenario, consistently with the collision model, the environment state is constant because, as we said,  each subsystem collides briefly with a sequence of sub-environments all initially in the same state. 
Of course one expects a strongly non markovian behavior if a given subsystem interacts repeatedly with {\em the same} sub-environment \cite{BP}.

The ME~(\ref{fin})  contains  both {\em local} Lindblad terms (i.e. Lindblad terms which act locally on the $m$-th carrier) and {\em two-body non local terms}  which couple the $M$ carrier with the $m' >m$.
More precisely  the $m$-th local  
term is the  super-operator
 \begin{eqnarray}
 \label{trieste2old1}
 {\cal L}_m (\cdots) &=& \frac{1}{2} \sum_{\ell,\ell'} {\gamma_m^{(\ell, \ell')}} \big[ 2 A_{S_m} ^{(\ell')} (\cdots)A_{S_m} ^{(\ell)}  \nonumber \\
 && -  A^{(\ell)}_{S_m}  A^{(\ell')}_{S_m}(\cdots)-  (\cdots)A^{(\ell)}_{S_m}  A^{(\ell')}_{S_m}   \big] \;,
 \end{eqnarray}
where the non negative matrix $\gamma^{(\ell,\ell')}_m$ is given by 
  \begin{eqnarray}\label{trieste3old}
 \gamma^{(\ell,\ell')}_m := \gamma \; 
\langle B_E^{(\ell)} B_E^{(\ell')} \; {\cal M}^{m-1}(\eta)\rangle_E\;,
 \end{eqnarray}
with $\gamma$ as in Eq.~(\ref{defgamma}).  Equation~(\ref{trieste3old}) defines the 
 correlation matrix of the sub-environment operators $B_E^{(\ell)}$ and $B_E^{(\ell')}$
  evaluated  (for the infinitesimal time interval $\Delta t$)
  on the density matrix ${\cal M}^{m-1}(\eta)$ which describes the state of the sub-environment after $m-1$ free evolution steps~\cite{NOTA1}.
For  $m'>m$ the cross terms of Eq.~(\ref{fin})  are defined instead as 
\begin{eqnarray}\label{d12}
{\cal D}^{(\rightarrow)}_{m,m'} (\cdots) &=& \sum_{\ell, \ell'} \gamma_{m,m'}^{(\ell, \ell')}  \;\;  A_{S_m}^{(\ell)}  \; \Big[ (\cdots) ,A_{S_{m'}}^{(\ell')} \Big]_- \nonumber \\
&-&  \sum_{\ell, \ell'} [\gamma_{m,m'}^{(\ell, \ell')}]^* \;  \;   \Big[(\cdots), A_{S_{m'}}^{(\ell')}  \Big]_- \;A_{S_m}^{(\ell)} \;
\end{eqnarray}
with $[\cdots,\cdots]_-$ being the commutation matrix and $\gamma^{(\ell,\ell')}_{m,m'}$ being the complex matrix~\cite{NOTA3}
   \begin{eqnarray} \label{gamma}
    \gamma^{(\ell,\ell')}_{m,m'}  :=   \gamma\; 
\langle B_{E}^{(\ell')} {\cal M}^{m'-m}( B_{E}^{(\ell)} \; {\cal M}^{m-1}(\eta)) \; \rangle_{E}\;.
 \end{eqnarray}
The coefficients $\gamma^{(\ell,\ell')}_{m,m'}$ introduce
 cross correlation among the carriers and depend  upon their {\em distance} $m'-m$. 
Furthermore, similarly to the the terms of Eq.~(\ref{trieste3old}), they  also depend on $m-1$ due to the fact that the model admits a {\em first} carrier.
 However if we assume that for large $m$ the sequence ${\cal M}^{m}(\eta)$ converges to 
   a final  point $\eta_0$, 
   then we can reach a stationary configuration where (for $m\gg1$)
$\gamma^{(\ell,\ell')}_{m,m'}$ only depends upon the distance $m' - m$ while $\gamma^{(\ell,\ell')}_{m}$ becomes constant in $m$,   i.e. 
   \begin{eqnarray}
   \gamma^{(\ell,\ell')}_{m,m'}  &\simeq&  \label{inf1}
\langle  B_{E}^{(\ell')}  {\cal M}^{m'-m}(  B_{E}^{(\ell)} \;\eta_0 )\rangle_{E}\;,\\
   \gamma^{(\ell,\ell')}_{m}  &\simeq& 
\langle B_{E}^{(\ell')}  \; B_{E}^{(\ell)} \;\eta_0 \rangle_{E}. \label{inf2}
 \end{eqnarray}
A similar behavior is obtained also if we assume $\eta$ to be a fix point for ${\cal M}$
(a reasonable hypothesis  if ${\cal E}$ is supposed to describe an environment
in its stationary configuration). In this case Eqs.~(\ref{inf1}), (\ref{inf2})  hold exactly 
for all $m$ and $m'$, 
with $\eta_0$ being replaced by $\eta$. Finally a case of particular interest is  the one in which
${\cal M}$ is the channel which sends every input state into $\eta$ (this is the extremal version of the last  two examples). Under this condition one expects that no correlations between the
various carriers can be established as the environmental sub-systems are immediately reset to their
initial state after each collision. Indeed in this case we have ${\cal M}(\theta) = \langle \theta\rangle_E \; \eta$
for all operators $\theta$, which, thanks to Eq.~(\ref{ASSUMPTION1}), yields $\gamma^{(\ell,\ell')}_{m,m'} =\gamma \langle B_{E}^{(\ell')} \eta \rangle_E \langle B_{E}^{(\ell)} \eta \rangle_E=0$ 
and hence ${\cal D}^{(\rightarrow)}_{m,m'}=0$.

\subsection{Correlations} 
Equation~(\ref{d12}) 
obeys to  proper time-ordering rules which guarantee that
the dynamical evolution of ${S}_m$ is {\em not} influenced by the  subsystems that follow it in the sequence, while it  might depend in a non trivial way on the carriers that precede it.
 Indeed when traced over the degree of freedom of the second carrier ${\cal S}_{m'}$, the cross term ${\cal D}^{(\rightarrow)}_{m,m'}$ nullifies, i.e. 
\begin{eqnarray} \label{prop1}
\left\langle {\cal D}^{(\rightarrow)}_{m,m'} (\cdots)\right\rangle_{S_{m'}} = 0 \;,
\end{eqnarray} 
while in general it does not disappear 
 when tracing over $S_{m}$ (it {\em does} disappear however if all the coefficients $\gamma^{(\ell,\ell')}_{m,m'}$ are real, see below). 
The evolution described by Eq.~(\ref{fin})  is thus {\em non-anticipatory}~\cite{GALL}, or 
in the jargon introduced in Ref.~\cite{VARI}, {\em semicausal} with respect to the ordering of the channels uses. To see this explicitly consider 
the  evolution of the reduced density matrix $\rho_{1,2}(t)$   of the first two carriers 
obtained by taking the partial trace 
of Eq.~(\ref{fin})
over all elements of ${\cal S}$ but $S_1$ and $S_2$. Noticing that $\langle {\cal L}_m(\cdots) \rangle_{S_m}=0$ and exploiting Eq.~(\ref{prop1}) we get 
 \begin{eqnarray}
{\dot \rho}_{1,2}(t) =  {\cal L}_1 (\rho_{1,2}(t) )   +
 {\cal L}_2 (\rho_{1,2}(t) )     + {\cal D}^{(\rightarrow)}_{1,2} (\rho_{1,2}(t) ) \;. \label{fin12}
 \end{eqnarray} 
 The resulting dynamics is  purely Markovian 
in full agreement   with the fact that $S_1, S_2$ couple weakly and sequentially with sub-environments ${\cal E}$ which have not interacted yet with other carriers. Tracing over $S_2$ we
can then derive the dynamical equation for $S_1$, i.e. 
${\dot \rho}_{1}(t) =  {\cal L}_1 (\rho_{1}(t) )$,
 which again is Markovian. 
Vice-versa the dynamics of $S_2$ cannot be expressed 
in terms of a close differential equation for $\rho_2(t)$ alone. Indeed
by 
taking the partial trace of Eq.~(\ref{fin12}) over $S_1$ we get 
 \begin{eqnarray} 
{\dot \rho}_{2}(t)& =&  
 {\cal L}_2 (\rho_{2}(t) )   \label{fin2} \\  \nonumber 
 &-& 2i \sum_{\ell,\ell'} \mbox{Im}[ \gamma_{1,2}^{(\ell,\ell')} 
] \; \left[ A_{S_2}^{\ell'}, \langle A_{S_1}(t), \rho_{1,2}(t)\rangle_{S_1}\right]_-\;, 
 \end{eqnarray} 
 where the last term explicitly depends upon the 
 joint density matrix of  $S_2$ {\em and} $S_1$~\cite{NOTA4}.
 This formally shows that in general  $S_1$ acts
 as controller for $S_2$, while  no back-action is allowed in the model. 

A case of special interest is represented by 
those situations in which the matrices $\gamma_{m,m'}^{(\ell,\ell')}$ are real. When this happens also 
the partial trace over $S_m$ of ${\cal D}^{(\rightarrow)}_{m,m'}$ nullifies, i.e. $\langle {\cal D}^{(\rightarrow)}_{m,m'} (\cdots)\rangle_{S_m}=0$. 
Accordingly the evolution of {\em any} subset of ${\cal S}$  is independent from the evolution of the remaining
carriers.   In this case hence our model becomes non-anticipatory with respect to all possible ordering of the carriers, 
describing hence a {\em non-signaling}  evolution~\cite{VARI} in which  the reduced
density matrix of each carrier evolves independently from the others. 
For instance in Eq.~(\ref{fin2}) the second line disappears yielding a Markovian equation also for $\rho_2(t)$, i.e. 
${\dot \rho}_{2}(t) =  {\cal L}_2 (\rho_{2}(t) )$.
\paragraph{Example:--}  As an application  we
focus on the  case in which the carriers and ${\cal E}$
form two sets of independent bosonic modes. In particular defining $a_m$ and $b_n$ to be
annihilation operators of the modes $S_m$ and $E_n$  respectively, we consider the  
 Hamiltonians $H_{S_m,E_n} = a_m \otimes b_n^\dag + a_m^\dag \otimes b_n$. We also
  take $\eta$ as the vacuum state of $E_n$. 
 and ${\cal M}$ as a lossy Bosonic quantum channel of transmissivity $\kappa$.
Notice that with these choices
 the Hermitian operators $A_{S_m}^{(\ell)}$ and $B_{E_n}^{(\ell)}$ entering in Eq.~(\ref{ggrss})
 are just quadrature operators of the fields, and that Eq.~(\ref{ASSUMPTION1}) is automatically
 verified for all $m$ since ${\cal M}(\eta)=\eta$.
The resulting model describes a correlated quantum channel 
  analogous to that of Ref.~\cite{LUPO} which mimics the transmission 
of  a sequence of optical pulses along an attenuating optical fiber  
characterized by finite relaxation times. 
The corresponding local ${\cal L}_m (\cdots)$ and cross term ${\cal D}^{(\rightarrow)}_{m,m'}$ 
entering in the final ME~(\ref{fin}) become respectively
 $\tfrac{\gamma}{2}  \left\{ 2 a_m (\cdots)a_{m} 
 -  a^\dag_{m}  a_{m}(\cdots)-  (\cdots)a^\dag_{m}  a_{m}  \right\}$
 and 
${\gamma} \kappa^{\tfrac{m'-m}{2}}
 \{ [ a_m (\cdots), a_{m'}^\dag]_-  - [  (\cdots)a_m^\dag, a_{m'}]_- \}$
  which exhibit an attenuation of the signals and an exponential decaying in the correlations
  (in particular ${\cal D}^{(\rightarrow)}_{m,m'} (\cdots)$ coincides with the cross term derived in Ref.~\cite{G}
for a collection of QED cavity modes coupled in cascade). 
 \section{Conclusions and perspectives}  
 In deriving the ME~(\ref{fin}) we assumed a specific ordering for the carriers of the model which
 implies that each elements in the sequence $S_1, S_2, \cdots, S_M$ can influence only the dynamical
 evolution of those which follow. This assumption was specifically introduced to account for the causal 
correlations that are present in many memory quantum channel models~\cite{GALL}. 
The collisional model however can be generalized to include more general correlations. 
For instance cyclical correlations can be accounted  by identifying $S_1$ 
with the $(M+1)$-th element of the set of carriers in such a way that $S_M$ can influence its dynamics. 
To do so it is sufficient to add  an independent set  ${\cal F}$ 
of sub-environments $F_1,F_2, \cdots, F_N$
which couple with ${\cal S}$ following a new ordering in which (say) all the carriers are shifted by 
one position (i.e. the element of ${\cal F}$ first interact with $S_2$, then with $S_3$, $S_4$, 
$\cdots$, $S_N$, and finally with $S_1$). A part from the new ordering the new couplings
are assumed to share the same properties of those that apply to ${\cal E}$ (in particular we require
that identities analogous to those in Eqs.~(\ref{ASSUMPTION1}), (\ref{defgamma})  hold). Under these
conditions (and assuming no direct interaction between ${\cal E}$ and ${\cal F}$) the
 ME~(\ref{fin}) will acquire new extra terms  which directly couple  each carrier 
with all the others. Specifically given  $m'>m$ we will have both a standard contribution 
of the form  ${\cal D}^{(\rightarrow)}_{m,m'}$ as in Eq.~(\ref{fin}) but also a  
contribution in which the role of $m$ and $m'$  are exchanged (i.e. something like ${\cal D}^{(\rightarrow)}_{m',m}$) that  originates from the couplings with ${\cal F}$. 
From this example it should be clear that by increasing the number sub-environmental sets and by properly tuning their interactions with ${\cal S}$ any sort of correlations can be built in dynamical
evolution of the system. 
 \newline
 
 \appendix
 \section{Technical sections} 
 
 In this section we  give the detailed derivation of Eq.~(\ref{fin}) and discuss its generalization to the case of non uniform collisional events. Subsequently we 
show how to include free evolution terms induced by local Hamiltonians operating on the carriers in the derivation of the ME.

\subsection{Derivation of Eq.~(\ref{fin})}

The starting point of the derivation is Eq.~(\ref{EQperR}) which under partial trace over ${\cal E}$ yields the identity 
  \begin{eqnarray}
&&\rho(n+1)=\rho(n) +  ( g\Delta t) \left\langle  {\cal C}^{\prime}_{{\cal S}, E_{n+1}}   \big(R(n)\otimes \eta\big) \right\rangle_{\cal E}   \label{EQperR111} \\
&& \qquad +  (g\Delta t)^2  \left\langle  {\cal C}^{\prime\prime}_{{\cal S}, E_{n+1}}  \big(R(n)\otimes \eta\big) \right\rangle_{\cal E} + {\cal O}\big((g\Delta t)^3\big)
\;,\nonumber 
\end{eqnarray} 
 In this expression  we need to specify the super-operators
${\cal C}^{\prime}_{{\cal S}, E_{n+1}}$ and ${\cal C}^{\prime\prime}_{{\cal S}, E_{n+1}}$ obtained by expanding 
${\cal C}_{{\cal S}, E_{n+1}}$ up to the second order in $g\Delta t$. To do so we notice that 
for each $m$ and $j$, the super-operators ${\cal U}_{S_{m},E_j}$ admit the following expansion,
\begin{eqnarray}
&&{\cal U}_{S_{m},E_j}  \label{compa}
= {\cal I}_{S_m,E_j} + (g\Delta t )\; {\cal U}_{S_{m},E_j}^\prime + (g\Delta t)^2 \; {\cal U}_{S_{m},E_j}^{\prime\prime} \nonumber \\
 && \qquad \qquad \qquad + {\cal O}\big((g\Delta t)^3\big)\;, 
 \end{eqnarray}
 with ${\cal I}_{S_m,E_j}$ being the identity map and with 
 \begin{eqnarray} 
 {\cal U}_{S_{m},E_j}^\prime (\cdots)&:=& - i \Big[ H_{S_m,E_j} , (\cdots) \Big]_-   \;, \\
 {\cal U}_{S_{m},E_j}^{\prime\prime} (\cdots)&:=&\nonumber  H_{S_m,E_j} (\cdots ) H_{S_m,E_j}\\
 && - \frac{1}{2} \Big[ H_{S_m,E_j}^2, (\cdots) \Big]_+ \;,  \label{compa1}
\end{eqnarray} 
where $[\cdots, \cdots]_-$ and $[\cdots, \cdots]_+$ represent the  commutator and the anti-commutator brackets respectively. 
From Eq.~(\ref{COLUMN}) it then follows that 
\begin{eqnarray}
 {\cal C}_{{\cal S}, E_j}^{\prime} &:=&
 \sum_{m=1}^M  {\cal M}_{ E_j}^{M-m+1} \circ {\cal U}_{S_{m},E_j}^\prime  
  \circ {\cal M}_{ E_j}^{m-1}\;, \label{COLUMN666} \\
   {\cal C}_{{\cal S}, E_j}^{\prime\prime} &:=&   {\cal C}_{{\cal S}, E_j}^{\prime\prime,a}   +    {\cal C}_{{\cal S}, E_j}^{\prime\prime,b}\;,
   \end{eqnarray}
   with 
   \begin{eqnarray}
 {\cal C}_{{\cal S}, E_j}^{\prime\prime,a} & := & \sum_{m=1}^M  {\cal M}_{ E_j}^{M-m+1} \circ {\cal U}_{S_{m},E_j}^{\prime\prime}   \circ {\cal M}_{ E_j}^{m-1}\;,   \nonumber \\
  {\cal C}_{{\cal S}, E_j}^{\prime\prime,b} &:= &
 \sum_{m' =m+1}^M  \sum_{m=1}^{M-1} {\cal M}_{ E_j}^{M-m'+1} \circ {\cal U}_{S_{m'},E_j}^{\prime}  
  \nonumber \\
  &&\qquad \quad \circ {\cal M}_{ E_j}^{m'-m}\circ {\cal U}_{S_{m},E_j}^{\prime}   \circ  {\cal M}_{ E_j}^{m-1}\;. \label{defCsec}
 \end{eqnarray} 
Replacing this into Eq.~(\ref{EQperR111}) we first notice that due to Eq.~(\ref{ASSUMPTION1}) the linear term in $g\Delta t$ nullifies. Indeed
we get 
\begin{widetext}
\begin{eqnarray}
 \left\langle  {\cal C}^{\prime}_{{\cal S}, E_{n+1}}   \big(R(n)\otimes \eta\big) \right\rangle_{\cal E} &=& 
-i   \sum_{m}  \left\langle  \Big[ H_{S_m,E_{n+1}},R(n)\otimes  {\cal M}_{ E_{n+1}}^{m-1} (\eta) \Big]_-\right\rangle_{\cal E} \nonumber \\
& = &
-i   \sum_{m}  \sum_{\ell} \left\langle  \Big[  A^{(\ell)}_{S_m}  \otimes B^{(\ell)}_{E_{n+1}},R(n)\otimes  {\cal M}_{ E_{n+1}}^{m-1} (\eta) \Big]_-\right\rangle_{\cal E} 
\nonumber \\
&=& -i   \sum_{m}  \sum_{\ell}   \Big[  A^{(\ell)}_{S_m} , \rho(n) \Big]_-  \left\langle B^{(\ell)}_{E_{n+1}} {\cal M}_{ E_{n+1}}^{m-1} (\eta) \right\rangle_{E_{n+1}} =0\;. 
 \end{eqnarray} 
  Vice-versa for the second order terms in $g\Delta t$ we get two contributions. The first is 
\begin{eqnarray}
 \left\langle  {\cal C}^{\prime\prime,a}_{{\cal S}, E_{n+1}}   \big(R(n)\otimes \eta\big) \right\rangle_{\cal E} &=& 
  \sum_{m}   \left\langle  \;  H_{S_m,E_{n+1}} (R(n)\otimes  {\cal M}_{ E_{n+1}}^{m-1} (\eta)) H_{S_m,E_{n+1}}
- \frac{1}{2}    \Big[ H_{S_m,E_{n+1}}, R(n)\otimes  {\cal M}_{ E_{n+1}}^{m-1} (\eta)\Big]_+ \right\rangle_{\cal E} 
\nonumber \\
& =&  \frac{1}{2}  \sum_{m} \sum_{\ell,\ell'} \langle B_E^{(\ell)} B_E^{(\ell')} \; {\cal M}^{m-1}(\eta)\rangle_E  \big[ 2 A_{S_m} ^{(\ell')} \rho(n) A_{S_m} ^{(\ell)}  -  A^{(\ell)}_{S_m}  A^{(\ell')}_{S_m} \rho(n)-  \rho(n) A^{(\ell)}_{S_m}  A^{(\ell')}_{S_m}   \big]  \nonumber 
\\ 
& =&   \frac{1}{\gamma} \; \sum_{m} {\cal L}_m (\rho(n)) \;,  \label{cprimeprimea}
 \end{eqnarray} 
 with ${\cal L}_m$  as in Eq.~(\ref{trieste2old1}). 
  The second term instead is 
\begin{eqnarray}
 \left\langle  {\cal C}^{\prime\prime,b}_{{\cal S}, E_{n+1}}   \big(R(n)\otimes \eta\big) \right\rangle_{\cal E} &=& 
 \sum_{m'>m} 
   \left\langle   {\cal U}_{S_{m'},E_{n+1}}^{\prime}  \circ {\cal M}_{ E_{n+1}}^{m'-m}\circ {\cal U}_{S_{m},E_{n+1}}^{\prime}      \big(R(n)\otimes 
   {\cal M}_{ E_{n+1}}^{m-1} (\eta)\big)
   \right\rangle_{\cal E}  \nonumber 
 \\ 
  &=& 
  -\sum_{m'>m} 
   \left\langle   \Big[  H_{S_{m'},E_{n+1}} , {\cal M}_{ E_{n+1}}^{m'-m}\Big(   \Big[  H_{S_{m},E_{n+1}} ,    R(n)\otimes 
   {\cal M}_{ E_{n+1}}^{m-1} (\eta) \Big]_- \Big) \Big]_-
   \right\rangle_{\cal E}  \nonumber \\
   &=&\sum_{m'>m}  \sum_{\ell, \ell'}  \Big\{ \langle B_{E}^{(\ell')} {\cal M}^{m'-m}( B_{E}^{(\ell)} \; {\cal M}^{m-1}(\eta)) \; \rangle_{E} \;\;  A_{S_m}^{(\ell)}  \; \Big[ \rho(n)  ,A_{S_{m'}}^{(\ell')} \Big]_- 
\nonumber \\
&&\qquad  \nonumber -  \langle B_{E}^{(\ell')} {\cal M}^{m'-m}( B_{E}^{(\ell)} \; {\cal M}^{m-1}(\eta)) \; \rangle_{E}^* \;  \;   \Big[\rho(n), A_{S_{m'}}^{(\ell')}  \Big]_- \;A_{S_m}^{(\ell)}  \Big\} \\
&& = \frac{1}{\gamma} \;\sum_{m'>m} \; {\cal D}^{(\rightarrow)}_{m,m'} (\rho(n)) \;,  \label{cprimeprimeb}
 \end{eqnarray} 
  with  ${\cal D}^{(\rightarrow)}_{m,m'}$ as in Eq.~(\ref{d12}).  
  Replacing all this into Eq.~(\ref{EQperR111})  gives 
 \begin{eqnarray}
 \frac{\rho(n+1)-  \rho(n)}{\Delta t}  =   \frac{g^2 \Delta t }{\gamma} \; \left\{  \sum_{m}  {\cal L}_m (\rho(n) )     
 + \sum_{m'>m} {\cal D}^{(\rightarrow)}_{m,m'} (\rho(n) )  \right\} 
\label{EQperR11245}   \;  + {\cal O}\big(g^3\Delta t^2\big) \;,
\end{eqnarray} 
 \end{widetext} 
which enforcing the limit~(\ref{defgamma}) yields the ME~(\ref{fin}). 

It is worth noticing that the above derivation 
still applies also if the collisional Hamiltonians (\ref{ggrss}) are not uniform. For instance suppose we have
\begin{eqnarray}\label{hamnonuni}
H_{S_mE_n} := \sum_\ell \; A^{(n,\ell)}_{S_m}  \otimes B^{(m,\ell)}_{E_n} \;,
\end{eqnarray} 
where now the operators acting on the carrier $S_m$ are allowed to explicitly depends upon 
the $n$ index which label the collisional events, and similarly the operators acting on the sub-enviroment are allowed
to explicitly depends upon the index $m$ which labels the carriers.
 Under these conditions one can verify that  Eq.~(\ref{EQperR11245}) still apply. 
In this case however, to account for the non uniformity of the couplings, 
the condition~(\ref{ASSUMPTION1}) needs to be generalized as follows
\begin{eqnarray}\label{ASSUMPTION2}
\left\langle B^{(m,\ell)}_{E} {\cal M}_{ E}^{m-1} (\eta) \right\rangle_{E} =0, \quad  \forall m, \ell \;.
\end{eqnarray} 
Furthermore both   ${\cal L}_m$ and ${\cal D}^{(\rightarrow)}_{m,m'}$ entering in Eq.~(\ref{EQperR11245}) become explicit functions of the carriers labels {\em and} of the
index $n$ which  plays the role of a temporal parameter for the reduced density matrix $\rho(n)$. Specifically 
the new super-operators are still  defined respectively as in Eqs.~(\ref{cprimeprimea}) and (\ref{cprimeprimeb})
with the operators  $A^{(n+1,\ell)}_{S_m}$ instead of $A^{(\ell)}_{S_m}$ and with
the coefficients $\langle B_E^{(\ell)} B_E^{(\ell')} \; {\cal M}^{m-1}(\eta)\rangle_E$ and 
$ \langle B_{E}^{(\ell')} {\cal M}^{m'-m}( B_{E}^{(\ell)} \; {\cal M}^{m-1}(\eta)) \; \rangle_{E}$ replaced by
$\langle B_E^{(m,\ell)} B_E^{(m,\ell')} \; {\cal M}^{m-1}(\eta)\rangle_E$ and 
$ \langle B_{E}^{(m',\ell')} {\cal M}^{m'-m}( B_{E}^{(m,\ell)} \; {\cal M}^{m-1}(\eta)) \; \rangle_{E}$ 
respectively.

The continuos limit~(\ref{defgamma}) can also still be defined by identifying $\lim_{\Delta t\rightarrow 0^+} A^{(n+1,\ell)}_{S_m}$ with the element $A^{(\ell)}_{S_m}(t)$ of a 
one parameter family of operators. As a result we get a 
 time-dependent  ME characterized by a Lindblad generator which explicitly depends
on $t$.

\subsection{Including local free evolution terms for the carriers}

Assume that between two consecutive collisions, the carriers undergo to a free-evolution described
by a (possibly time-depedent)  Hamiltonian $H_{{\cal S}}(t): =\sum_{m} h_{S_m}(t)$ which are local (i.e. no direct interactions between the carriers is allowed).
Under these conditions Eq.~(\ref{fin}) still holds in the proper interaction picture representation at the price of
allowing the generators of the ME to be explicitly time dependent. 

To see this we first notice that under the assumption that the collision time $\Delta t$  is much shorter than the time interval that elapses between two consecutive
collisional events  (i.e. $\Delta t \ll \tau_{n}-\tau_{n-1}$),
the unitary operator which describes the evolution of the $m$-th carrier in its interaction with ${\cal E}$  is now given by 
\begin{eqnarray}\label{hh2222}
&&U_{S_m {\cal E}}^{(n)}
 :=  U_{S_m,E_n} V_{S_m}( \tau_n, \tau_{n-1})\;U_{S_m,E_{n-1}} \\ \nonumber
 &&\;\cdots \; V_{S_m}(\tau_2,\tau_1)\;U_{S_m,E_2} \; V_{S_m}(\tau_1,0) \;U_{S_m,E_1},
\end{eqnarray} 
where $U_{S_m,E_n}$ are the collisional transformations, $\tau_n$ is the time at which the $n$-th collision takes place, and where 
$V_{S_m}(\tau_{n},\tau_{n-1}):={\cal T}
 \exp[-i \int_{\tau_{n-1}}^{\tau_n} dt' h_{S_m}(t')]$ is the 
unitary operator which describes the free-evolution of $S_m$ between the $(n-1)$-th and the $n$-th collision
(in this expression ${\cal T}
 \exp[\cdots]$ indicates the time-ordered exponential which we insert to explicitly account for possibility that  the $h_{S_m}$ will be 
  time-dependent). 
Define hence the operators 
\begin{eqnarray}
\bar{A}^{(n,\ell)}_{S_m}  : =  V_{S_m}^\dag (\tau_n,0)  \;  {A}^{(\ell)}_{S_m}\;   V_{S_n} (\tau_n,0)\;,
\end{eqnarray} and
the
Hamiltonian 
\begin{eqnarray} \label{fffs}
\bar{H}_{S_m,E_n} &:=& 
 V_{S_n}^\dag (\tau_n,0) \; {H}_{S_m,E_n} V_{S_n}  (\tau_n,0)
\nonumber \\ &=& \sum_\ell \; \bar{A}^{(n,\ell)}_{S_m} \otimes B^{(\ell)}_{E_n} 
\;,
\end{eqnarray}
which describes  the coupling between $S_m$ and ${\cal E}$ in the interaction representation associated with the free evolution of $S_m$.
Notice that the  operators $\bar{A}^{(n,\ell)}_{S_m}$ are explicit  functions of the index $n$ which labels the collisions as in the case of Eq.~(\ref{hamnonuni}) (here
however the terms operating on ${\cal E}$ are kept uniform). 
Observing that  for all $\ell$  one has $V_{S_m}(\tau_\ell,\tau_{\ell-1}) V_{S_m}(\tau_{\ell-1},\tau_{\ell-2}) = V_{S_m}(\tau_\ell,\tau_{\ell-2})$
we  can now write Eq.~(\ref{hh2222}) as
\begin{eqnarray}\label{hh22223}
&&U_{S_m {\cal E}}^{(n)}
 := V_{S_m}(\tau_{n},0) \;  \bar{U}_{S_m,{\cal E}}^{(n)},
\end{eqnarray} 
where $\bar{U}_{S_m,{\cal E}}^{(n)}$ is the unitary that defines the collisions of $S_m$  with the sub-environments in the interaction representation, i.e.  
\begin{eqnarray}\label{hh2222555}
&&\bar{U}_{S_m,{\cal E}}^{(n)}
 :=  \bar{U}_{S_m,E_n} \;\bar{U}_{S_m,E_{n-1}}\cdots  \bar{U}_{S_m,E_1},
\end{eqnarray} 
with 
\begin{eqnarray} 
 \bar{U}_{S_m,E_n}  = \exp[ - i g\;  \bar{H}_{S_m,E_n} \Delta t ]\;.
 \end{eqnarray} 
 Similarly we can express   the super-operators ${\cal W}^{(n,M)}$  as 
 \begin{eqnarray}
{\cal W}^{(n,M)}  &=&   {\cal V}_{{\cal S}}(\tau_{n},0) \circ \; \bar{\cal W}^{(n,M)} \;, \\
\bar{\cal W}^{(n,M)} &:=&   \bar{\cal C}_{{\cal S}, E_n}  \circ \cdots  \circ \bar{\cal C}_{{\cal S}, E_1}\;, \\
 \bar{\cal C}_{{\cal S}, E_j} &:=& {\cal M}_{ E_j} \circ \bar{\cal U}_{S_{M},E_j} \circ \cdots
   \circ {\cal M}_{ E_j}
 \circ \bar{\cal U}_{S_{1},E_j},
\end{eqnarray}
with ${\cal V}_{{\cal S}}(\tau_{n},0)$ being the super-operator associated with the
joint free unitary evolution   obtained by combining all the local terms  of the carriers, i.e. 
$V_{\cal S}(\tau_{n},0) := V_{S_1}(\tau_{n},0) \cdots V_{S_M}(\tau_{n},0)$.
 Defining hence $\bar{R}(n)$ the state of ${\cal S}$ and of the first elements of ${\cal E}$ after $n$ collisions in the interaction representation induced
by $V_{\cal S}(\tau_{n},0)$ as 
\begin{eqnarray} 
\bar{R}(n) =V_{\cal S}^\dag(\tau_n,0) \; R(n) \; V_{\cal S}(\tau_n,0)\;,
\end{eqnarray} 
we get a recursive expression analogous to Eq.~(\ref{rec}) with ${\cal C}_{{\cal S}, E_{n+1}}$ replaced by 
$\bar{\cal C}_{{\cal S}, E_{n+1}}$, i.e. 
\begin{eqnarray}
\bar{R}(n+1)&=&
\bar{\cal C}_{{\cal S}, E_{n+1}}  (\bar{R}(n)\otimes \eta)\;. \label{rec1}
\end{eqnarray} 
More precisely this expression formally coincides with that which, as in the case described at the end of the previous section, one would have obtained starting from a collisional model
in which no free evolution of the carriers is allowed but the collisional events are not uniform. Indeed the generators of the dynamics $\bar{H}_{S_m,E_n}$ do have 
the same form of the Hamiltonians~(\ref{hamnonuni}). Following the same prescription given there, 
we can then get an expression for the reduced density matrix 
 $\bar{\rho}(n)=\langle \bar{R}(n) \rangle_{\cal E}$  which represents the state of the carriers after $n$ collisions in the interaction picture with respect to the free evolution 
 generated by $H_{\cal S}(t)$.
Enforcing the limit~(\ref{defgamma}) under the condition~(\ref{ASSUMPTION2}),  one can verify that  $\bar{\rho}(t)$ obeys 
to a ME analogous to Eq.~(\ref{fin}) with the operators $A_{S_m} ^{(\ell')}$ being replaced by 
the time-dependent operators $\bar{A}^{(\ell)}_{S_m}(t):= \lim_{\Delta t\rightarrow 0^+} \bar{A}^{(n,\ell)}_{S_m}$. 
\newline 

VG acknowledges support by the FIRB-IDEAS 
project (RBID08B3FM).


\begin{thebibliography}{99}
\bibitem{NOTA1}
Equation~(\ref{trieste2old1}) can   also be casted in a more traditional form~\cite{LIND,PET}  by diagonalizing  the matrix $\gamma^{(\ell,\ell')}_m$:
  this allows one to identify the decay rates  of the system with the non-negative eigenvalues $r^{(\ell)}_{m}$ of $\gamma^{(\ell,\ell')}_m$ and the
associated Lindblad operators $L_{S_m}^{(\ell)}$  with a proper linear combinations of the  $A_{S_m}^{(\ell)}$.
\bibitem{NOTA3}  Eq.~(\ref{gamma}) can be written as  
$\gamma \langle {\cal M}_{\text{H}}^{m'-m}( B_{E}^{(\ell')})   B_{E}^{(\ell)}  {\cal M}^{m-1}(\eta) \rangle_{E}$ where ${\cal M}_{\text{H}}$ is 
 the   Heisenberg adjoint  of  ${\cal M}$. 
 In this form  $\gamma^{(\ell,\ell')}_{m,m'}$ appears to  be 
 a generalized two-time correlation function of the environment operators $B_{E}^{(\ell')}$ and $ B_{E}^{(\ell)}$ with respect to the density operator ${\cal M}^{m-1}(\eta)$.
\bibitem{NOTA4}
 An approximate equation for $\rho_{2}(t)$ alone can be obtained
  by requiring  $\rho_{1,2}(t)\simeq \rho_{1}(t) \otimes \rho_{2}(t)$ for each time $t$
(this is equivalent to consider $S_1$ as part of an effective environment which is weakly coupled to $S_2$). In this case Eq.~(\ref{fin2}) becomes
$\dot{\rho}_{2}(t) \simeq  - i [ H_2^{(eff)}(t) , \rho_{2}(t)]_- + {\cal L}_2 (\rho_{2}(t) )$,
with $H_2^{(eff)}(t):=2  \sum_{\ell,\ell'} \text{Im}[\gamma_{1,2}^{(\ell,\ell')}]\; \langle A_{S_1}^{(\ell)} \rho_{1}(t) \rangle_{S_1} A_{S_2}^{(\ell')}$
 being an effective time dependent Hamiltonian of $S_2$.
\end{thebibliography}
\end{document}